\documentclass[reprint,amsmath,amssymb, aps,prl,floatfix,superscriptaddress]{revtex4-1}
\usepackage[dvipsnames]{xcolor}
\usepackage{graphicx} 
\usepackage[separate-uncertainty=true]{siunitx}
\usepackage[version=4]{mhchem}
\usepackage{comment}
\usepackage{tikz}
\usepackage{scalerel}
\usepackage{xifthen}
\usepackage{chngcntr}

\DeclareSIUnit\molar{M}
\DeclareSIUnit{\wtpc}{wt\%}

\usetikzlibrary{svg.path}
\usepackage[colorlinks,allcolors=blue]{hyperref}

\newcommand{\figref}[2][{}]{Fig.\ \ref{#2}\ifthenelse{\isempty{#1}}{}{\,(#1)}}
\renewcommand{\vec}{\mathbf}

\makeatletter
\def\maketitle{
\@author@finish
\title@column\titleblock@produce
\suppressfloats[t]}
\makeatother

\newcommand{\beginsupplement
}{%
    \setcounter{table}{0}
    \renewcommand{\thetable}{S\arabic{table}}%
    \setcounter{figure}{0}
    \renewcommand{\thefigure}{S\arabic{figure}}%
    \setcounter{equation}{0}
    \renewcommand{\theequation}{S\arabic{equation}}%
    \setcounter{section}{0}
    \setcounter{subsection}{0}
    \renewcommand{\thesection}{\arabic{section}}
    \maketitle
     }

\definecolor{orcidlogocol}{HTML}{A6CE39}
\tikzset{
  orcidlogo/.pic={
    \fill[orcidlogocol] svg{M256,128c0,70.7-57.3,128-128,128C57.3,256,0,198.7,0,128C0,57.3,57.3,0,128,0C198.7,0,256,57.3,256,128z};
    \fill[white] svg{M86.3,186.2H70.9V79.1h15.4v48.4V186.2z}
                 svg{M108.9,79.1h41.6c39.6,0,57,28.3,57,53.6c0,27.5-21.5,53.6-56.8,53.6h-41.8V79.1z M124.3,172.4h24.5c34.9,0,42.9-26.5,42.9-39.7c0-21.5-13.7-39.7-43.7-39.7h-23.7V172.4z}
                 svg{M88.7,56.8c0,5.5-4.5,10.1-10.1,10.1c-5.6,0-10.1-4.6-10.1-10.1c0-5.6,4.5-10.1,10.1-10.1C84.2,46.7,88.7,51.3,88.7,56.8z};
  }
}

\newcommand\orcid[1]{\href{https://orcid.org/#1}{\mbox{\scalerel*{
\begin{tikzpicture}[yscale=-1,transform shape]
\pic{orcidlogo};
\end{tikzpicture}
}{|}}}}

\begin{document}

\setlength{\unitlength}{1cm}

\newcommand{\goeaffila}{Max Planck Institute for Dynamics and Self-Organization, Am Fa\ss{}berg 17, 37077 G\"ottingen, Germany}
\newcommand{\goeaffilb}{Institute for the Dynamics of Complex Systems, Georg August Universit\"at G\"ottingen, Germany}
\newcommand{\twaffil}{Physics of Fluids Group, Max Planck Center for Complex Fluid Dynamics and J. M. Burgers Center for Fluid Dynamics, University of Twente, PO Box 217,7500AE Enschede, Netherlands}
\newcommand{\hydaffil}{Department of Mechanical and Aerospace Engineering, Indian Institute of Technology Hyderabad, Kandi, Sangareddy, Telengana- 502285, India}

\author{Carola M. Buness\orcid{0000-0003-0437-0538}} 
\affiliation{\goeaffila}
\affiliation{\goeaffilb}

\author{Avi Rana} 
\affiliation{\hydaffil}

\author{Corinna C. Maass\orcid{0000-0001-6287-4107}}
\email{c.c.maass@utwente.nl}
\affiliation{\twaffil}
\affiliation{\goeaffila}

\author{Ranabir Dey\orcid{0000-0002-0514-7357}}
\email{ranabir@mae.iith.ac.in}
\affiliation{\hydaffil}
\affiliation{\goeaffila}

\begin{abstract}
Ciliated microswimmers and flagellated bacteria alter their swimming trajectories to follow the direction of an applied electric field exhibiting electrotaxis.
Both for matters of application and physical modelling, it is instructive to study such behaviour in synthetic swimmers.
We show here that under an external electric field, self-propelling active droplets autonomously modify their swimming trajectories in microchannels, even undergoing `U-turns', to exhibit robust electrotaxis. 
Depending on the relative initial orientations of the microswimmer and the external electric field, the active droplet can also navigate upstream of an external flow following a centre-line motion, instead of the oscillatory upstream trajectory observed in absence of electric field.
Using a hydrodynamic theory model, we show that the electrically induced angular velocity and electrophoretic effects, along with the microswimmer motility and its hydrodynamic interactions with the microchannel walls, play crucial roles in dictating the electrotactic trajectories and dynamics. 
Specifically, the transformation in the trajectories during upstream swimming against an external flow under an electric field can be understood as a reverse Hopf bifurcation for a dynamical system.
Our study provides a simple methodology and a systematic understanding of manoeuvring active droplets in microconfinements for micro-robotic applications especially in biotechnology. 
\end{abstract}

\newcommand{\mytitle}{Electrotaxis of self-propelling artificial swimmers in microchannels}
\title{\mytitle} 

\maketitle

\section*{Introduction}
Biological microswimmers, e.g. bacteria, algae, paramecium, sperm cells, adapt their swimming strategies in response to external cues resulting in directed motion exemplified by the various \textit{taxes}, like chemotaxis in response to chemical concentration gradients \cite{berg1972_chemotaxis,berg1975_chemotaxis,karmakar2021_state}, phototaxis in response to light gradients \cite{jekely2009_evolution}, rheotaxis in response to external flows \cite{bretherton1961_rheotaxis,conrad2018_confined,mathijssen2019_oscillatory}, and gravitaxis in response to gravitational field \cite{sengupta2023_planktonic}. 
Interestingly, both flagellated and ciliated microswimmers, like \textit{E. coli} \cite{shi1996_effect,liu1999_electrokinetic}, \textit{Paramecium} \cite{machemer-rohnisch1996_electricfield,ogawa2006_physical}, \textit{T. pyriformis} \cite{kim2013_galvanotactic}, and \textit{C. hirtus} \cite{daul2022_galvanotaxis}, also exhibit a directed swimming in response to external electric fields leading to galvanotaxis or electrotaxis (also see \cite{green2023_how}).
The mechanism for electrotaxis under an applied DC electric field is dependent on the cell surface structure for bacteria \cite{shi1996_effect}, and on the augmentation and reversal of the ciliary beating pattern for ciliates like \textit{Paramecium} \cite{ogawa2006_physical}.
The latter is classically referred to as the Ludloff phenomenon \cite{ludloff1895_untersuchungen}. 
In the context of self-propelling synthetic microswimmers, it is worth noting that mimicry of such biological electrotactic behaviour can have far reaching consequences for their control and manoeuvrability during technological applications, like targeted cargo (drug) delivery \cite{wang2012_nano,patra2013intelligent}.\\

Self-propelling artificial microswimmers can be broadly classified into two groups- one, active colloids, like Janus particles, which swim by initiating self-diffusio-, or self-electro-phoresis of a chemical species triggered by catalytic reactions at the interface initiated by a `fuel' chemical \cite{zottl2016_emergent,liebchen2021_interactions,zottl2023_modeling}; two, active droplets, which swim predominantly using interfacial tension gradient  created by spontaneous symmetry breaking of a `fuel' surfactant at the droplet interface~\cite{toyota2009_selfpropelled,thutupalli2011_swarming,peddireddy2012_solubilization,herminghaus2014_interfacial,maass2016_swimming,birrer2022_we,dwivedi2022_selfpropelled,michelin2023_selfpropulsion}.
It is fascinating to note that both types of artificial microswimmers adapt their swimming dynamics in response to external cues, and mimic some of the taxes exhibited by biological microswimmers, like chemotaxis \cite{sundararajan2008_catalytic,jin2017_chemotaxis,hokmabad2022_chemotactic}, rheotaxis \cite{uspal2015_rheotaxis,katuri2018_crossstream,dwivedi2021_rheotaxis,dey2022_oscillatory,stark2016swimming}, and gravitaxis \cite{palacci2010_sedimentation,castonguay2023_gravitational,stark2016swimming}.
However, investigations into electrotaxis of artificial microswimmers remain scarce and incomplete.\\

The rotational and translational dynamics of self-propelling platinum-polystyrene Janus particles under an external DC electric field were experimentally demonstrated in the context of manoeuvrability of active colloids using solid boundaries \cite{das2015_boundaries}.
The inherent asymmetry in the surface zeta potential of the Janus particles resulted in a rotation which aligned the swimming direction with the applied electric field \cite{das2015_boundaries}. 
Subsequent theoretical analysis \cite{bayati2019_electrophoresis} revealed that the angular velocity responsible for this realignment of the active Janus particles is dependent on the asymmetric electrokinetic effects, and the strengths of the surface activity and the applied electric field. 
Recently, a combination of AC and DC electric fields in a 3D microelectrode geometry were also used to demonstrate controllability of catalytic Janus rods \cite{guo2018electric}.
However, the response of isotropic active swimmers lacking inherent asymmetry, e.g. active droplets, to external electric fields remains unexplored. 
This cannot be predicted \textit{a priori} based on the response of asymmetric Janus particles, which also swim in the diffusion dominated regime unlike isotropic autophoretic microswimmers which need finite advective perturbations for the spontaneous symmetry breaking of the `fuel' surfactant. 
Furthermore, the limited available studies on the electrotaxis of self-propelling artificial microswimmers remains restricted to unbounded \cite{bayati2019_electrophoresis} or open surface \cite{das2015_boundaries,guo2018electric} geometries.
Keeping in mind that real life applications demanding electrical manoeuvrability of artificial microswimmers will invariably involve strong confinements, the effects of confining walls in dictating the dynamics in response to electrical cues need to be studied carefully.\\

Here, we study the swimming dynamics of self-propelling active droplets in microchannels in response to external DC electric fields.
Under an opposing electric field, the  active droplet reorients its swimming direction by 180$^\circ$ to preferentially swim along the direction of the applied field, and thereby exhibiting electrotaxis.
During this sharp `U'-turn, the active droplet also switches its location from the vicinity of one side wall of the microchannel to the other. 
Since in a majority of applications the microswimmers also need to navigate external flows, we further study the adaptation of the swimming strategy during upstream rheotaxis in response to external DC electric fields.
Intriguingly, under an electric field,  the active droplet switches from the oscillatory upstream rheotactic trajectory \cite{dey2022_oscillatory} to a linear upstream trajectory along the channel center-line.
We explain the observed dynamics using an approximate hydrodynamic model which incorporates the electrical effects- an electrical force on the swimmer, electrophoresis, electroosmosis, and an electrically induced angular velocity, and the interaction of the microswimmer with the microchannel walls.
Our model successfully describes all the phenomenological observations, including revealing the electro-rheotactic behaviour of the active droplets as a reverse Hopf bifurcation of a dynamical system.  
We believe that our study of the electrotaxis of self-propelling droplet microswimmers in narrow confinements, also in presence of an external flow, offers a simple way for controlling and manoeuvring synthetic isotropic microswimmers for a wide range of applications, including biomedical.\\  

\section*{Results}
\begin{figure*}
    \centering
    \includegraphics[width=\textwidth]{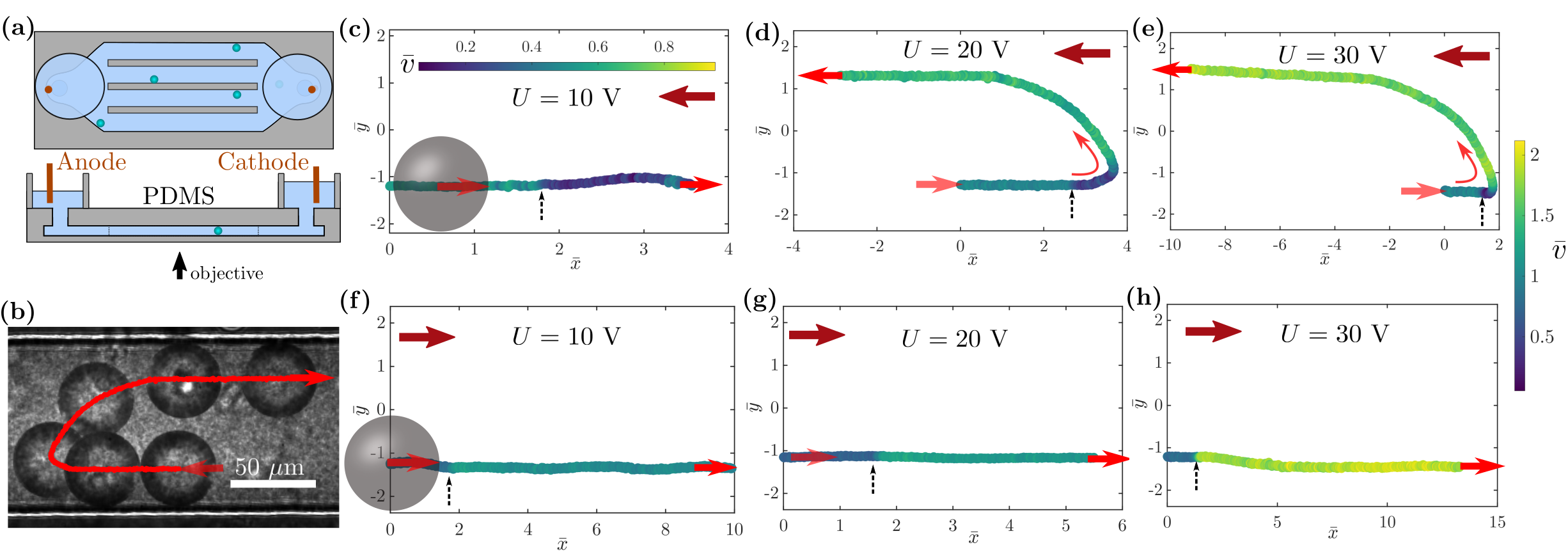}
    \caption{\textbf{Electrotaxis of an active droplet in a microchannel with increasing  electrical voltage.} \textbf{(a)} Schematic of the microfluidic setup, having microchannels of height $h \approx 52$ $\mu$m and width $2w \approx 100$ $\mu$m, used for microscopy. The electrical voltage $U$ is applied between the two electrodes dipped into the two reservoirs. \textbf{(b)} Time-lapse super-position from fluorescence microscopy, showing an active droplet (diameter $2R_d \approx 42$ $\mu$m) changing its initial ($U=0$ V) swimming trajectory (to the left along the bottom wall) by making a `U-turn' to swim (along the top wall) in the direction of the applied electric field $\mathbf{E}$ (to the right). \textbf{(c)-(e)} Trajectory of the active droplet, colour coded by the non-dimensional instantaneous velocity $\bar{v}$, for increasing value of $U$, when $\mathbf{E}$ (maroon arrow) is opposite to the initial swimming direction of the droplet (light red arrow), and \textbf{(f)-(h)} when $\mathbf{E}$ is oriented along the initial swimming direction. The vertical black arrow marks the  location when the electrical voltage is switched on.}
    \label{fig:electrotaxis}
\end{figure*}

\begin{figure}
    \centering
    \includegraphics[width=\columnwidth]{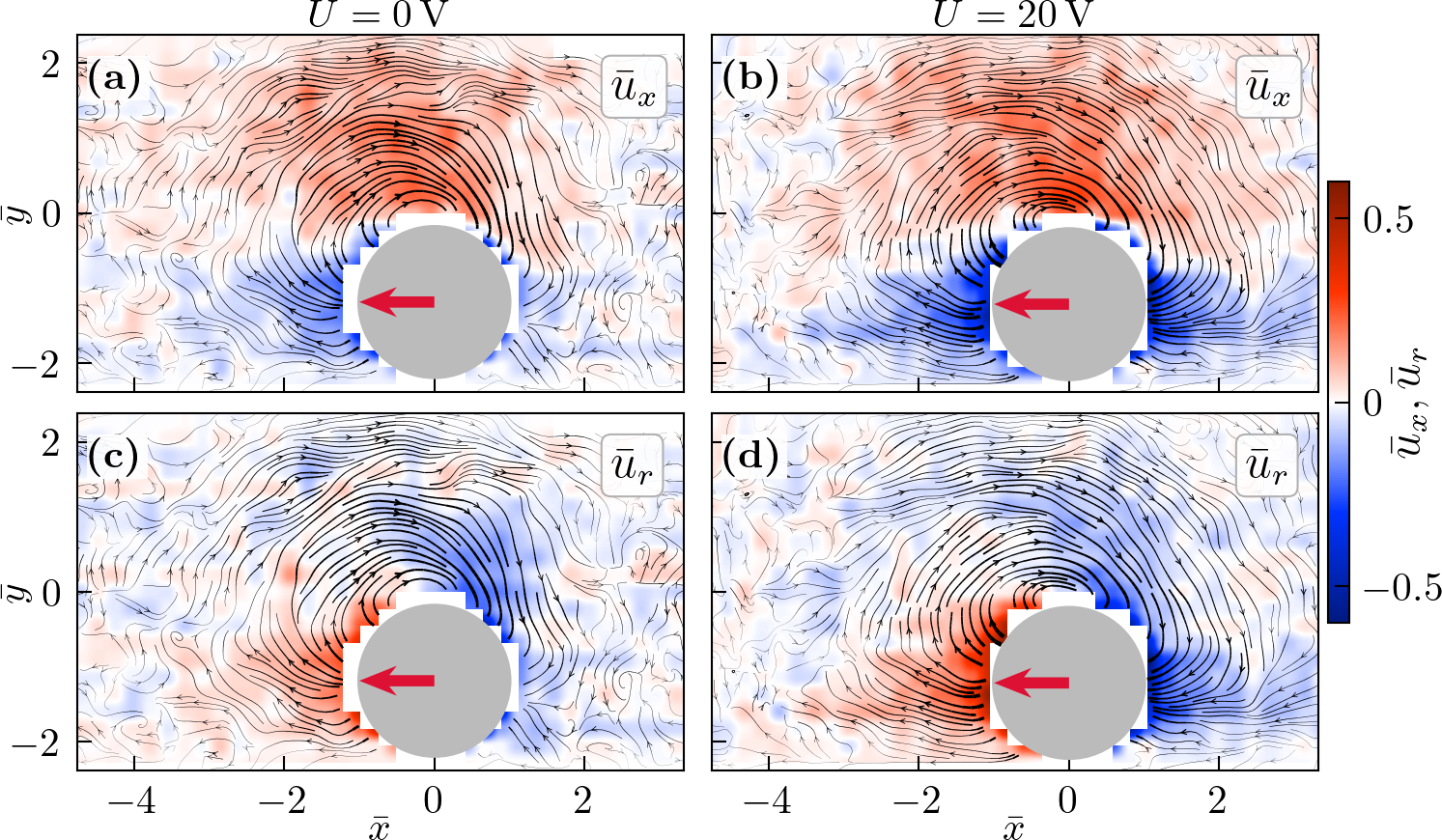}
    \caption{\textbf{Velocity field generated by the active droplet during negative electrotaxis.} Distribution of the local velocity components $\bar{u}_x$ and $\bar{u}_r$ (colourmap) and the corresponding streamlines for the flow generated by the active droplet during its self-propulsion without any electrical voltage ($U=0$ V; left column \textbf{(a)} and \textbf{(c)}), and during electrotaxis along $\mathbf{E}$ ($U=20$ V; right column \textbf{(b)} and \textbf{(d)}).}
    \label{fig:velocity field}
\end{figure}

We use a well-established experimental system of active CB15 oil droplets~\cite{herminghaus2014_interfacial,michelin2023_selfpropulsion} solubilising~\cite{peddireddy2012_solubilization} in a supramicellar solution of the cationic surfactant TTAB. 
We modified an  experimental assay (Fig.~\ref{fig:electrotaxis} (a)) for rheotaxis of active droplets in microchannels~\cite{dey2022_oscillatory} to investigate electrotaxis by inserting electrodes into the microfluidic chip outlets and applying an electrical voltage $U$. 
This creates a linear electric field parallel to the channel axis- $\mathbf{E}=\pm U/L \mathbf{\hat{x}}$. As $L$ is fixed, $|\vec{E}|$ is directly proportional to $U$.
We studied the response of the active droplets to the imposed electrical voltage between \SI{0}{\volt} and \SI{30}{\volt} (Fig.~\ref{fig:electrotaxis} (b)-(h)), analysing droplet trajectories and flow fields (Fig.~\ref{fig:velocity field}) via video microscopy tracking and Particle Image Velocimetry (PIV). 
For details, see the Supplementary Material. 
To study electro-rheotactic effects (Fig.~\ref{fig:electro-rheotaxis}), an optional pressure driven flow was imposed via the chip inlet.

We non-dimensionalize all spatial variables by the active droplet radius $R_d\approx21$ $\mu$m, velocities by the inherent swimming velocity in a quiescent medium $v_0 \approx 29.5$ $\mu$m$\cdot s^{-1}$, and time, as well as angular velocities, by $R_d/v_0$ and its inverse.
All non-dimensionalized variables are represented by overbars.
The instantaneous swimming velocity of the active droplet in the quasi-2D microchannel is given by $\mathbf{v}=v\mathbf{\hat{e}}=v\left(\cos \psi \mathbf{\hat{x}} + \sin \psi \mathbf{\hat{y}}\right)$, where $\mathbf{\hat{e}}$ is the unit vector defining the swimming orientation, and $\psi$ is the orientation angle which is positive when measured in an anti-clockwise sense relative to $+\mathbf{\hat{x}}$.\\  

\subsection *{Electrotaxis of active droplets}
With no electric field applied, and in a quiescent swimming medium, droplets reliably attach to the walls of microfluidic geometries~\cite{dey2022_oscillatory,jin2019_fine}. 

When $\mathbf{E}$ is applied opposite to the swimming direction of the active droplet, one of two things can happen -- one, when $U\leq 10$ V, the droplet continues to swim following its original trajectory along a wall, i.e. opposite to the direction of $\mathbf{E}$, but with a reduced instantaneous velocity $\bar{v}$ (Fig.~\ref{fig:electrotaxis}(c)).  
Two, when $U$ is higher, $\bar{v}$ initially reduces. Subsequently, the active droplet simultaneously rotates and translates altering its trajectory to swim in the direction of $\mathbf{E}$, along the opposite wall (Fig.~\ref{fig:electrotaxis}(d)-(e)). 
This is referred to here as (negative) electrotaxis, in accordance with the classical terminology used for electrotaxis of microorganisms \cite{ogawa2006_physical}. 
During this `U-turn' under $\mathbf{E}$, $\bar{v}$ increases compared to its intrinsic self-propulsion velocity along a microchannel wall without electrical effects (Fig.~\ref{fig:electrotaxis} (d) or (e)). 
When $\mathbf{E}$ is applied along the initial swimming direction of the active droplet, it only exhibits electrotaxis with $\bar{v}$ increasing with  $U$ (Fig.~\ref{fig:electrotaxis} (f)-(h)).\\
Next, we investigate the difference in the hydrodynamic signature comparing the velocity field generated by the active droplet during electrotaxis with that generated during its inherent self-propulsion (Fig.~\ref{fig:velocity field}). 
As the droplet microswimmer swims along the direction of $\mathbf{E}$, the axial component $(\bar{u}_x)$  of the local velocity field is higher both at the anterior and posterior ends of the microswimmer compared to the case without $\mathbf{E}$ (compare Figs.~\ref{fig:velocity field} (a) and (b)). 
We also evaluate the radial component of the velocity field $(\bar{u}_r)$ considering the droplet centroid as the origin (Figs.~\ref{fig:velocity field} (c)).
Importantly, $\bar{u}_r$ is higher both at the anterior and posterior halves of the microswimmer during electrotaxis (compare Figs.~\ref{fig:velocity field} (c) and (d)).
Furthermore, $\bar{u}_r$ decays over longer radial distance from the droplet centroid.
The higher magnitude and longer decay of $\bar{u}_r$ over the anterior and posterior regions of the active droplet, and the overall flattening of the streamlines, are due to the additional stokeslet~\cite{kim2005_microhydrodynamics} like velocity field component during electrotaxis, superimposed on its inherent velocity field. 
\subsection*{Electro-rheotaxis of active droplets}
\begin{figure}
    \centering
    \includegraphics[width=\columnwidth]{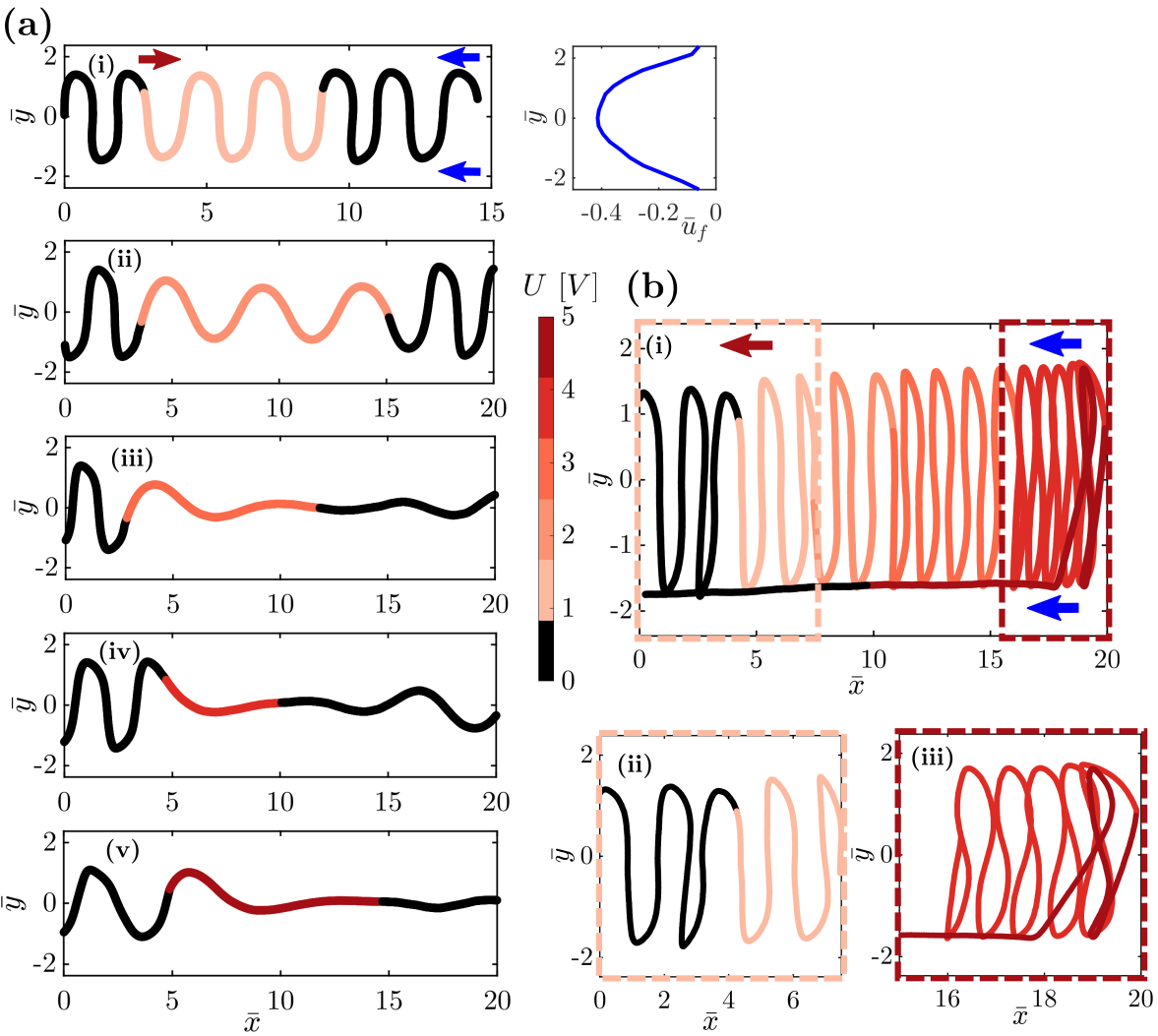}
    \caption{\textbf{Electro-rheotaxis of an active droplet.} Variation in the trajectory of the active droplet undergoing oscillatory upstream rheotaxis, against a pressure-driven flow in a microchannel, on application of an electric field (maroon arrow) with increasing strength oriented \textit{opposite} to the imposed flow from right (blue arrows) \textbf{(a)(i)-(v)}, and oriented \textit{along} the imposed flow \textbf{(b)(i)-(iii)}. The colourbar represents the applied electrical voltage $(U)$, with black corresponding to $U=0$ V. The imposed flow profile is shown in the right panel of (a)(i). Coloured boxes in (b)(ii) and (b)(iii) show blow-ups of data ranges in b(i).}
    \label{fig:electro-rheotaxis}
\end{figure}
Active droplets can navigate external flows in microchannels by exhibiting an oscillatory upstream swimming trajectory, i.e. positive rheotaxis~\cite{dey2022_oscillatory}, reminiscent of many biological microswimmers~\cite{uppaluri2012_flow,mathijssen2019_oscillatory}.
Here, we investigate the relationship between such rheotactic and the above described electrotactic swimming of the active droplets.
For all the experiments, the droplet microswimmer is always initially oriented upstream (along $+\mathbf{\hat{x}}$) of the imposed pressure-driven flow (always along $-\mathbf{\hat{x}}$).  
For low $U$ and $\mathbf{E}$ applied opposite to the direction of the imposed flow (i.e. along $+\mathbf{\hat{x}}$), the active droplet apparently continues to exhibit upstream oscillatory rheotaxis against the flow (Fig. ~\ref{fig:electro-rheotaxis}(a)(i)).
However, with increasing $U$, the amplitude of oscillation during upstream rheotaxis decreases, while the wavelength progressively increases (Fig.~\ref{fig:electro-rheotaxis}(a)(i)-(iii)).
Eventually, interestingly, the droplet microswimmer swims upstream of the imposed flow only by following a linear trajectory along the channel center-line (Fig.~\ref{fig:electro-rheotaxis}(a)(iv), (v)). \\

When $\mathbf{E}$ is applied along the direction of the imposed pressure-driven flow (i.e. along $-\mathbf{\hat{x}}$), the active droplet continues to exhibit upstream oscillatory rheotaxis even for relatively higher $U$, as compared to the case when $\mathbf{E}$ is applied opposite to the flow (Fig.~\ref{fig:electro-rheotaxis}(b)(i)).
However, with increasing $U$, the shape of the oscillatory trajectory gradually changes specifically near the microchannel walls, and for this case, the wavelength of oscillation progressively decreases (Fig.~\ref{fig:electro-rheotaxis}(b)(i) and (ii)).
Eventually, for high $U$, the droplet reorients to swim downstream of the imposed flow following a linear trajectory along one of the walls of the microchannel (Fig.~\ref{fig:electro-rheotaxis}(b)(i) and (iii)).\\

\subsection*{A hydrodynamic model for electrotaxis and electro-rheotaxis}
\begin{figure}
    \centering
    \includegraphics[width=\columnwidth]{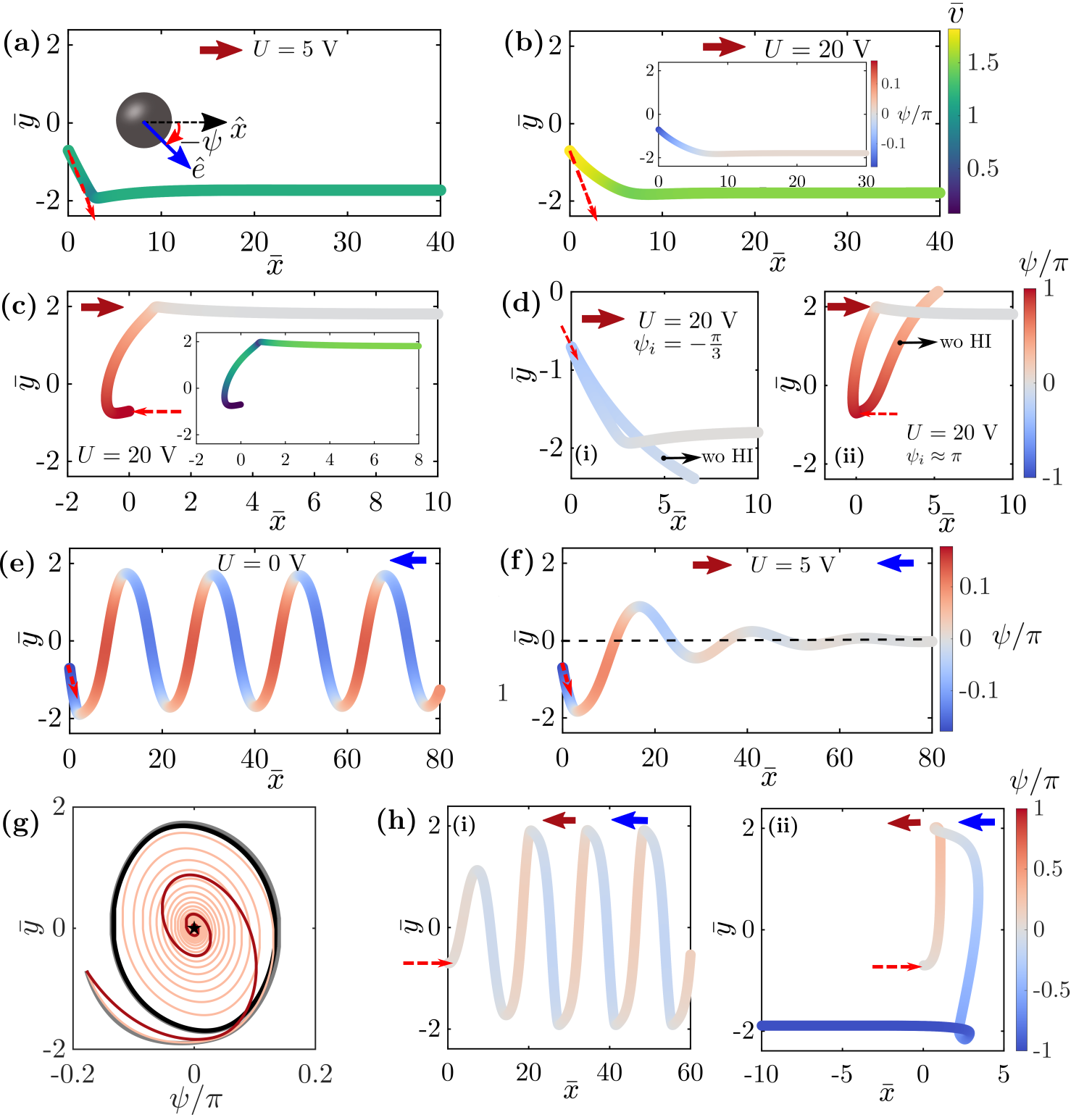}
    \caption{\textbf{Electrotactic dynamics of an active droplet predicted by a hydrodynamic model (Eqs. \ref{eqn:xdot}-\ref{eqn:psidot}).} Electrotaxis of the active droplet when the initial orientation $(\psi_i)$ of the microswimmer (red dotted line) is within $\pm \pi/2$ relative to the direction of the applied electric field $\mathbf{E}$ (along $+\hat{x}$) for \textbf{(a)} $U=5$ V and \textbf{(b)} $U=20$ V. \textbf{(c)} Electrotactic dynamics when $\psi_i = \pi$, \textit{opposite} to the direction of $\mathbf{E}$. \textbf{(d)} Electrotactic dynamics with and without hydrodynamic interactions (wo HI) of the microswimmer with the microchannel walls for different values of $\psi_i$ and $U$. \textbf{(e,f)} Theoretical predictions of the swimming trajectory and orientation $\psi$ of the microswimmer during upstream rheotaxis against a pressure-driven flow (from right; blue arrows) with an average flow velocity of $\bar{u}_f=0.23$ when \textbf{(e)} $U=0$, and \textbf{(f)} when $U=5$ V is applied along the upstream direction of the flow (maroon arrow). \textbf{(g)} Trajectories in the $(\psi,\bar{y})$ phase space for different values of $U$-- $0$ (grey), $1$ V (cream), $5$ V (dark red), with $\bar{u}_f=0.23$. \textbf{(h)} Theoretical predictions of the electro-rheotactic dynamics with increasing $U$ when $\mathbf{E}$ is oriented downstream and the microswimmer is initially oriented upstream of the flow.}
    \label{fig:theo_model}
\end{figure}
To understand the electrotactic and the electro-rheotactic dynamics we use a 2D hydrodynamic model commensurate with the quasi-2D experimental setup.
The velocity field generated by the self-propelling active droplet is approximated by an equivalent pusher-type squirmer (centroid: $\bar{x}_0$, $\bar{y}_0$) velocity field
, comprising of a force-dipole (non-dimensional strength $\bar{\beta}$) and a source-dipole (non-dimensional strength $\bar{\gamma}$) \cite{lighthill1952_squirming,spagnolie2012_hydrodynamics,de2019flow,kuron2019_hydrodynamic}, as 
$\bar{\beta}\mathbf{\bar{u}_{fd}}+\bar{\gamma}\mathbf{\bar{u}_{sd}}$.
The active droplet is positively charged because of the interfacial adsorption of the cationic surfactant (TTAB) monomers.   
Thus, under $\mathbf{E}$, the charged active droplet experiences an electric force which, in the thin electric double layer (or the Smoluchowski) limit, is given by $\mathbf{F_e}=6\pi \epsilon R_d \zeta_s \mathbf{E}= 6\pi \epsilon R_d \zeta_s U/L \mathbf{\hat{x}}$ \cite{kim2005_microhydrodynamics}. 
Here, for the theoretical model the uniform $\mathbf{E}$ is along $+\mathbf{\hat{x}}$ unless otherwise mentioned, $\epsilon$ is the permittivity of the aqueous surfactant solution, and $\zeta_s \sim O(10)$ mV is the surface zeta potential of the active droplet. 
The velocity field solely due to $\mathbf{F_e}$ is approximated by superposing a stokeslet flow field $\mathbf{\bar{u}_{st}}$ \cite{kim2005_microhydrodynamics,spagnolie2012_hydrodynamics} of strength $\bar{\alpha} \sim \frac{F_e}{8\pi\mu v_o R_d}$, 
in line with the experimentally observed flow fields in Fig.~\ref{fig:velocity field} (b,d), as discussed above.
The hydrodynamic interaction of the self-propelling droplet with the `no-slip' microchannel walls at $\bar{y}=\pm\bar{w}$ is addressed by image systems for each of the stokeslet $(\bar{\alpha} \mathbf{\bar{u}^*_{st}})$, force-dipole $(\bar{\beta} \mathbf{\bar{u}^*_{fd}})$, and source-dipole $(\bar{\gamma} \mathbf{\bar{u}^*_{sd}})$ flow fields \cite{blake1971_note,spagnolie2012_hydrodynamics,graaf2016_understanding,de2019flow,radhakrishnan2023_confinement}.
The translational dynamics under $\mathbf{E}$ is finally obtained using Fax\'en's first law \cite{kim2005_microhydrodynamics,spagnolie2012_hydrodynamics,graaf2016_understanding, kuron2019_hydrodynamic,radhakrishnan2023_confinement} for a spherical squirmer with \textit{non-zero} net hydrodynamic force in an ambient velocity field given by 
\begin{equation}\bar{\beta}\mathbf{\bar{u}_{fd}}+\bar{\gamma}\mathbf{\bar{u}_{sd}}
+\bar{\alpha}\mathbf{\bar{u}_{st}}+\mathbf{\bar{u}_f}+\Sigma_{walls} \left(\bar{\alpha}\mathbf{\bar{u}^*_{st}}+\bar{\beta}\mathbf{\bar{u}^*_{fd}}+\bar{\gamma}\mathbf{\bar{u}^*_{sd}}\right).\label{eqn:allflow}\end{equation} Here, $\Sigma$ represents the image systems for both channel walls, and $\mathbf{\bar{u}_f}$ represents any externally imposed flow velocity field. 

The resulting equations of motion can be written as
\begin{equation}
    \begin{split}
    \dot{\bar{x}}_0&=\cos{\psi}+\bar{u}_{eo}+\bar{u}_{ep}-\frac{3}{2} \bar{u}_f^a\left[1-\left(\frac{\bar{y}_0}{\bar{w}}\right)^2-\frac{1}{3\bar{w}^2}\right]\\
    &+ \bar{\alpha}\left[\mp \frac{3}{4(\bar{y}_0 \pm \bar{w})}\pm \frac{1}{32(\bar{y}_0 \pm \bar{w})^3}\right] \\
    &+\bar{\beta} \left[\pm \frac{3}{8(\bar{y}_0 \pm \bar{w})^2} \pm \frac{0.078125}{(\bar{y}_0 \pm \bar{w})^4}\right] \sin{2\psi}\\
    &+\bar{\gamma} \left[\mp \frac{1}{4(\bar{y}_0 \pm \bar{w})^3} \pm \frac{3}{16(\bar{y}_0 \pm \bar{w})^5}\right] \cos{\psi}\\
    \end{split}
    \label{eqn:xdot}
\end{equation}
\begin{equation}
    \begin{split}
      \dot{\bar{y}}_0&=\sin{\psi}+\bar{\beta}\left[\mp \frac{3(1-3\sin^2{\Psi})}{8(\bar{y}_0 \pm \bar{w})^2} \pm \frac{(7-11\sin^2{\Psi})}{64(\bar{y}_0 \pm \bar{w})^4}\right]\\
    &+\bar{\gamma} \left[\mp \frac{1}{(\bar{y}_0 \pm \bar{w})^3} \pm \frac{1}{8(\bar{y}_0 \pm \bar{w})^5}\right] \sin{\psi}\\
    \end{split}
    \label{eqn:ydot}
\end{equation}
In Eq.\ \ref{eqn:xdot}, $\mathbf{\bar{u}_{eo}}=\frac{1}{v_0}\mathcal{M}_{eo} \mathbf{E}$ is the contribution from the concomitant plug-like electroosmotic flow of the aqueous ionic surfactant solution under $\mathbf{E}$, with the electroosmotic mobility $\mathcal{M}_{eo} \approx -4.4 \times 10^{-9}$ m$^2$/V$\cdot$s; see~Supplementary Fig.\ \ref{sifig:eocalibration}), and $\mathbf{\bar{u}_{ep}}=\frac{1}{v_0}\mathcal{M}_{ep}\mathbf{E}$ is the additional electrophoretic velocity of the charged squirmer in response to $\mathbf{F}_e$, with the associated mobility $\mathcal{M}_{ep}=\frac{\epsilon \zeta_s} {\mu} \sim O(10^{-9})$ m$^2$/V$\cdot$s  \cite{kim2005_microhydrodynamics}. The fourth term in Eq.\ \ref{eqn:xdot} is the contribution from the optional imposed pressure-driven (plane Poiseuille type) flow along $-\mathbf{\hat{x}}$  with a non-dimensional average velocity $\bar{u}_f^a$ (=0 or 0.23). 

The temporal change in the swimming orientation $\psi$ of the charged squirmer is dictated by two effects. 
First, induced by the ambient velocity field as given by $\mathbf{\dot{\hat{e}}}=\mathbf{\bar{\Omega}}\times\mathbf{\hat{e}}$, where \[\mathbf{\bar{\Omega}}=\frac{1}{2} \nabla \times \left[\mathbf{\bar{u}_f}+\Sigma_{walls} \left(\bar{\alpha}\mathbf{\bar{u}^*_{st}}+\bar{\beta}\mathbf{\bar{u}^*_{fd}}+\bar{\gamma}\mathbf{\bar{u}^*_{sd}}\right)\right]\] following Fax\'en's second law \cite{kim2005_microhydrodynamics,spagnolie2012_hydrodynamics,graaf2016_understanding, kuron2019_hydrodynamic,radhakrishnan2023_confinement}.
Second, the change in $\psi$ due to an additional angular velocity $\mathbf{\bar{\Omega}_e}=\bar{\omega} (\mathbf{\hat{e}} \times \mathbf{\hat{x}})$ which tends to align $\mathbf{\hat{e}}$ with the direction of $\mathbf{E}$.
This is analogous to the angular velocity theoretically estimated for active Janus particles under an electric field \cite{bayati2019_electrophoresis}.
For isotropic active droplets, $\mathbf{\bar{\Omega}_e}$ stems from the electric response of the charged surface in presence of the activity at the interface (flux of TTAB monomers) on application of $\mathbf{E}$, when $\mathbf{\hat{e}}$ is not aligned with $\mathbf{E}$. 
Here, we approximate $\bar{\omega}=\Gamma |\mathbf{E}| \sim O(10^{-1})$ following the theoretical estimates in \cite{bayati2019_electrophoresis}, where the parameter $\Gamma$ is dependent on $\zeta_s$ and activity.
The equation for $\psi$ can be written as:
\begin{equation}
\begin{split}
\dot{\psi}&=-\frac{3}{2} \bar{u}_f^a \frac{\bar{y}_0} {\bar{w}^2}-\bar{\omega}\sin{\psi} \mp \bar{\beta} \frac{3}{16(\bar{y}_0 \pm \bar{w})^3} \sin{2\psi}\\
&\pm \bar{\gamma} \frac{3}{8(\bar{y}_0 \pm \bar{w})^4} \cos{\psi}\\
\end{split}
 \label{eqn:psidot}
\end{equation}
In Eqs. \ref{eqn:xdot}-\ref{eqn:psidot}, each term with $\pm/\mp$ represents two terms due to the image systems at the bottom (first sign) and top (second sign) walls. 
Eqs. \ref{eqn:ydot} and \ref{eqn:psidot} are solved numerically for $\bar{y}_0(t)$ and $\psi(t)$, and subsequently, Eq. \ref{eqn:xdot} is numerically integrated to obtain $\bar{x}_0(t)$. \\

\section*{Discussion}
In Fig.\ \ref{fig:theo_model} we illustrate the predictions of the model for a selection of initial conditions $(\bar{y}_{0}^i, \psi_i)$, external fields $\vec{E}$ and $\bar{u}_f^a$.
Panels (a) and (b) compare similar $\bar{y}_{0}^i = -0.7$, $\psi_i = -60 ^\circ$  under the respective applied voltages of 5  and \SI{20}{\volt}.
Driven by the electrically induced $\mathbf{\bar{\Omega}_e}$,  $\psi$ rotates towards a steady-state value of $\psi=0$ (inset in (b)) from any initial forward angle $-\pi/2<\psi_i<\pi/2$ (red dotted arrows). 
Notably, the final wall distance decreases with increasing $U$ (see similar trends in the experimental trajectories in Fig.\ \ref{fig:electrotaxis} g,h).
The steady state speed $\bar{v}$ increases with increasing $U$ because of the contribution from $\mathbf{\bar{u}_{ep}}$ with positive $\mathcal{M}_{ep}$ (compare Figs. \ref{fig:theo_model}(a,b) and the experimental data in Fig. \ref{fig:electrotaxis}(f-h)).

Panel (c) uses the same $\bar{y}_{0}^i = -0.7$, but now starting from an orientation of $\psi_i\approx\pi$, i.e. opposite to the direction of $\vec{E}$. 
Driven by $\mathbf{\bar{\Omega}_e}$ and $\mathbf{\bar{u}_{ep}}$, the droplet migrates cross-stream, and after a `U-turn', again continues to exhibit electrotaxis in the direction of $\vec{E}$. 
This reflects the experimental behaviour shown in Fig.\ \ref{fig:electrotaxis} (d,e).

The interaction with the channel walls has a significant influence on the aforementioned reorientation dynamics. 
We demonstrate this by setting the coefficients for the starred flow fields in Eqn.\ \ref{eqn:allflow} to zero: in consequence, all terms proportional to $\bar{\alpha}$, $\bar{\beta}$ and $\bar{\gamma}$ in Eqns.\ \ref{eqn:xdot} -- \ref{eqn:psidot} will be dropped.
Panel (d) shows the evolution from the forward and backward initial orientation plotted in (b) and (c): without these hydrodynamic wall interactions, the droplet crashes into the wall, while their inclusion replicates the steady-state wall alignment we observe in experiments.

Panels (e)--(g) investigate the effect of an imposed Poiseuille flow in the direction of $-\hat{\vec{x}}$. In (e), the squirmer is initially oriented upstream, $|\psi_i|<\pi/2$, and $U=0$.
The squirmer undergoes an oscillatory upstream rheotaxis mainly due the variation in the angular velocity of the total ambient flow field~\cite{dey2022_oscillatory}: see also the black experimental trajectories in Fig. \ref{fig:electro-rheotaxis}(a). This oscillatory trajectory in the absence of $\mathbf{E}$ represents a stable limit cycle in the $(\psi,\bar{y})$ phase space (black line in Fig. \ref{fig:theo_model}(g)).

\begin{figure}[h!]
    \centering
    \includegraphics[width=\columnwidth]{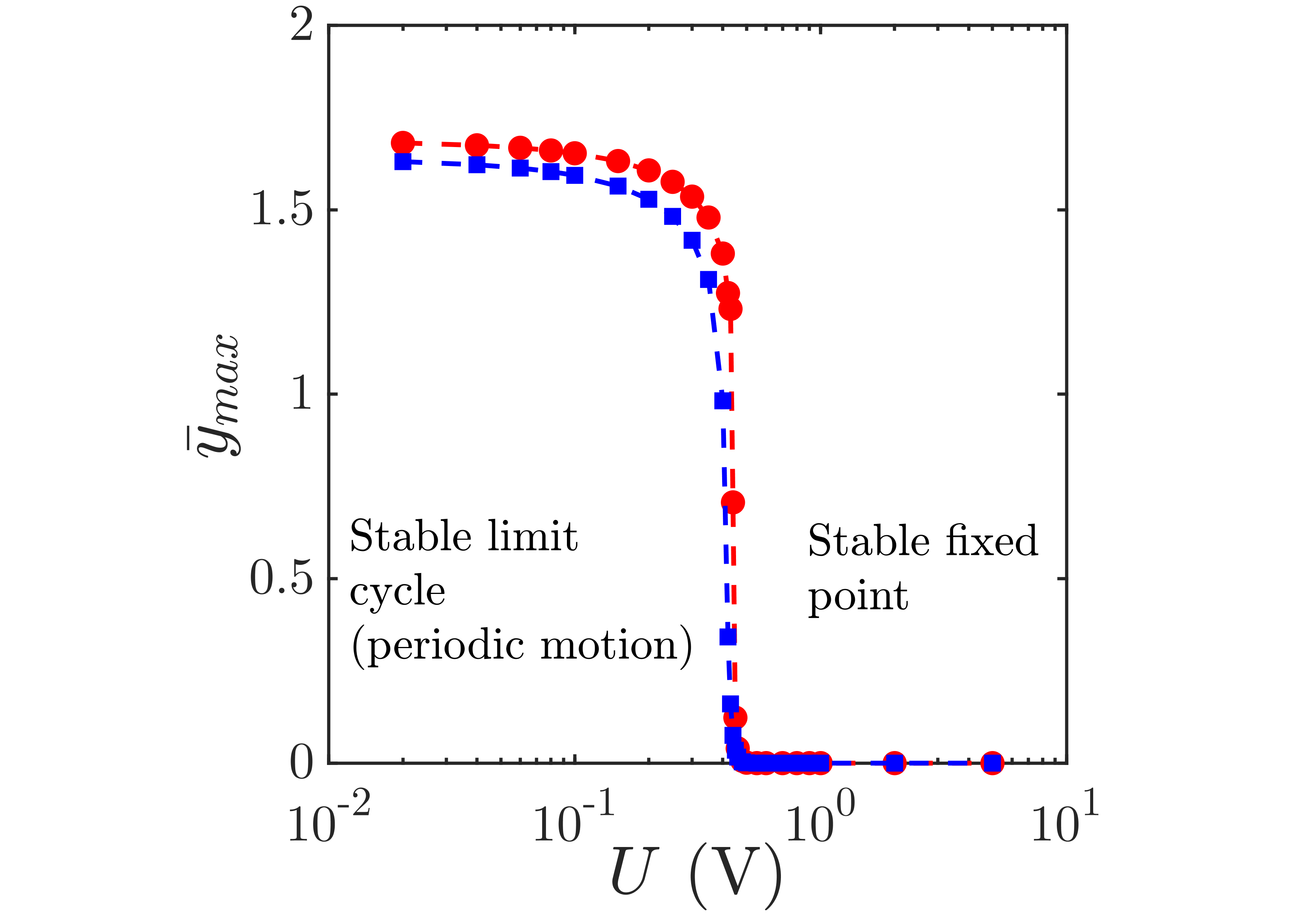}
    \caption{A bifurcation diagram showing the variation of the amplitude of the steady-state oscillatory trajectory during electro-rheotaxis with the applied electrical voltage. Variations for two different flow rates with non-dimensional average flow velocities- $\bar{u}_f^a$=0.23 (red markers) and 0.5 (blue markers) are shown here.}
    \label{sifig:bifurcation}
\end{figure}

On application of $\mathbf{E}\propto+\hat{\vec{x}}$, the squirmer transitions to a stable trajectory along the channel-centerline, as seen in experiment (compare Fig.\ref{fig:theo_model}(f) and Fig. \ref{fig:electro-rheotaxis}(a)(iii)-(v)). Above a critical $U_{c}$ depending on $\bar{u}_f^a$, a stable fixed point emerges at $\psi=\bar{y}=0$, and trajectories starting from any initial upstream condition $(\psi_i,\bar{y}_0^i)$ spiral towards it (cream and dark red curves in Fig. \ref{fig:theo_model}(g)). Under these conditions, the electro-rheotactic dynamics undergo the reverse of the classical Hopf bifurcation with increasing $U$ and $U_c\approx\SIrange{400}{500}{\milli\volt}$ (suppl. Fig.\ \ref{sifig:bifurcation}). The decay of the oscillatory trajectory towards a stable fixed point is also evident in the experimental trajectories for $U\geq\SIrange{1}{2}{\volt}$ (Fig. \ref{fig:electro-rheotaxis}(a)), which exhibit a decrease in the amplitude and a progressive increase in the wavelength. The decay was not observable for lower voltages, where the channel length is insufficient to capture the very gradual transition. 

Finally, in panel (h), we apply the model to the case where both $\vec{E}$ and $\vec{u_f}$ are applied in negative $-\vec{\hat{x}}$, with an initial condition of $\psi_i=0, \bar{y}_{0}^i=-0.7$. 
For low voltage, the droplet oscillates upstream of the imposed flow. 
For high $U$, $\mathbf{\bar{\Omega}_e}$ and $\mathbf{\bar{u}_{ep}}$ are strong enough to reorient the squirmer downstream, but now close to the wall.
These correspond to the experimental behaviour shown in Fig.\ \ref{fig:electro-rheotaxis} (b).

\subsection*{Summary and Outlook}
We have systematically studied the effects of electric fields and external flow on a generic experimental model swimmer in a narrow confinement. 
We found a variety of behaviours depending on initial conditions, specifically the relative orientations of the microswimmer and the applied electric field, including electrotaxis, `U-turns', tunable upstream oscillation, center-line motion, and wall attachment.
These at first sight disorganised dynamics can be fully explained by a deterministic hydrodynamic model consisting of ingredients like the squirmer motility, hydrodynamic interaction with the confinement boundaries, electroosmotic and electrophoretic effects, electrically induced angular velocity, and the imposed flow profile. 
We show that the electrically induced angular velocity which tends to align the instantaneous swimming orientation of the charged microswimmer with the direction of the electric field, along with the hydrodynamic interactions with the microchannel walls, play crucial roles in turning the microswimmer initiating electrotaxis.
We find that the combined effects of the external flow and the electric field result in a unique center-line swimming of the active droplet, instead of a channel-wide oscillation observed in presence of only the external flow.
Using phase portraits, we illustrate that this transformation in the trajectories represents a reverse Hopf bifurcation for a dynamical system.

Previously, only the electrical response of asymmetric active Janus particles were shown ~\cite{das2015_boundaries,guo2018electric} and theoretically explained in an unbounded quiescent domain ~\cite{bayati2019_electrophoresis}. 
Our study establishes the generality of the electrotactic behaviour of self-propelling artificial microswimmers, by demonstrating and explaining the phenomenon for isotropic autophoretic microswimmers in narrow confinements, both in absence and presence of an external flow.
Furthermore, the fact that the electrically modulated swimming behaviour of active droplets can be described by a general hydrodynamic model shows that self-propelling droplets are a surprisingly robust model system, whose behaviour can be compared with and extrapolated to other synthetic and biological electrophoretic microswimmers.

The electrotactic behaviour of the droplet microswimmers, including the `U-turn', under an electric field is apparently identical to the observations for microorganisms ~\cite{kim2013_galvanotactic}.
Hence, our study demonstrates a precise imitation of yet another kind of biological taxis by synthetic microswimmers, albeit with different underlying `sensory' mechanisms.

Electro-rheotaxis offers a simple way for manoeuvring active agents in complex geometries by just exploiting their inherent surface charge due to the adsorption of cationic surfactants -- controlling wall attachment, centerline motion, oscillation, translation etc. in a highly robust manner, paving the way for possible micro-robotic applications ~\cite{michelin2023_selfpropulsion}.

Finally, since the application of an electric field alters the orientation and long-range hydrodynamic signature of active droplets, electric fields can be also used to tune their mutual interactions and collective behaviour.
These can be the scope of interesting future studies.

\par
\section*{Acknowledgements}
\par
We thank Anirban Jana and Alexandra Alexiu for helping with the preliminary experiments during their summer internships. RD also thanks Anupam Gupta and Harish N. Dixit for insightful discussions.
RD acknowledges SERB, Department of Science and Technology, Government of India for partial support through grant no. SRG/2021/000892.

\bibliography{electrotaxis}

\cleardoublepage
\title{\mytitle{}: Supplementary Materials}
\beginsupplement


\section{Microfluidic fabrication}
\begin{figure}[h!]
    \centering
    \includegraphics[width=\columnwidth]{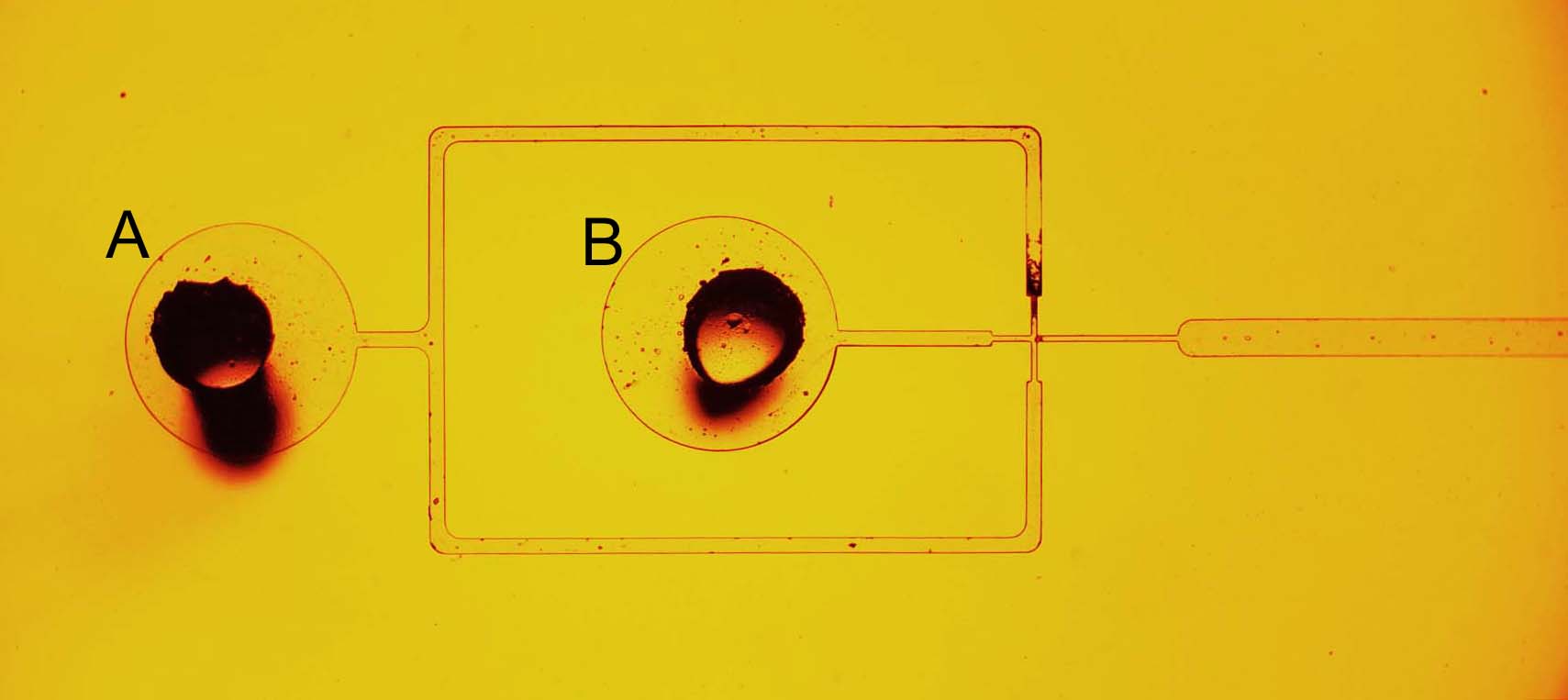}
    \caption{Droplet production in a PDMS cross junction}
    \label{sifig:dropletproduction}
\end{figure}

\subsection*{Droplet generation}
PDMS chips for all experimental geometries were fabricated in house using standard soft lithography techniques~\cite{qin2010_soft}.
We produced monodisperse oil-in-water emulsions in PDMS flow focussing junctions using the protocol described in~
\cite{dey2022_oscillatory} (Fig.\ \ref{sifig:dropletproduction}. The inner phase consisted of  4-cyano-40-(2-methylbutyl)-biphenyl (CB15, Synthon Chemicals), the outer phase of an aqueous solution of tetradecytrimethyl ammonium bromide (TTAB, Sigma Aldrich) at \SI{0.1}{\wtpc}.

\subsection*{Channel setup}
We performed the experiments in a quasi-two-dimensional PDMS structure containing 12 parallel channels of length \SI{9}{\mm}, width \SI{100}{\um} and height \SI{52}{\um} (see \ref{sifig:channel_sketch}b), corresponding to one droplet diameter in height and two in width. To ensure identical boundary conditions at all channel walls, the structure was capped with a PDMS coated glass slide at the bottom, and bonded via plasma treatment, following the method outlined in \cite{dey2022_oscillatory}. Two hollow plastic cylinders were subsequently attached with PDMS above the inlet and outlet of the structure (both punched in from above), providing fluid reservoirs as shown in Fig. \ref{sifig:channel_sketch}c).
\section{Experiments}
\subsection*{Generation of electric fields in microchannels}
\begin{figure}
\noindent\includegraphics[width=\columnwidth]{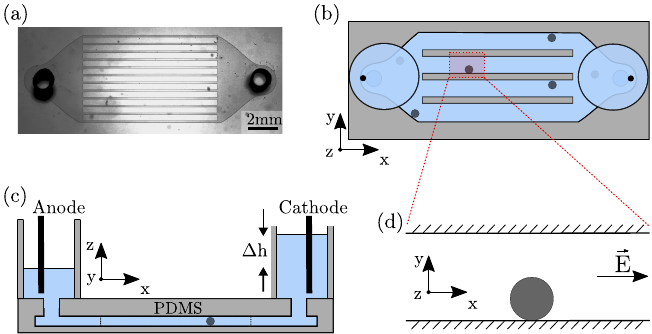}
 \caption{\label{sifig:channel_sketch} Sketch of the channel setup. (a) micrograph, top view, (b) top view schematic, (c) side view schematic with reservoirs  and electrodes, (d) Definitions of coordinate and field directions, top view. }
\end{figure}

We placed two titanium wire electrodes (length $\approx\SI{2}{\cm}$, diameter
$\SI{250}{\um}$), each with one end  in one of the reservoirs,
extending \SIrange{2}{3}{\mm} into the surfactant solution  at a distance in $x$ of \SI{1.5\pm 0.1}{\cm}, as shown in Fig.\ \ref{sifig:channel_sketch} (a,c). The electric field was applied either at constant set voltage (with an
error of \SI{\pm 0.15}{\volt}), or using a square wave AC signal. Since the observation area ($<500$x\SI{500}{\um^2}) was small compared to the electrode distance and measurements were recorded in the middle channels of the geometry in Fig.\ \ref{sifig:channel_sketch} (a), we assume the field to be with good accuracy homogeneous and $\parallel x$.

\subsection*{Pressure driven flow}
We generated a pressure driven flow along the microchannels by filling the reservoirs up to levels differing by a height $\Delta h$ on the order of 100 micrometers. As $\Delta h$ levels out over long times, only short time experiments were taken, with a duration $<$ 240 \SI{}{\second}.  From PIV data recorded before and after one such experiment, we found $\Delta v/v\approx$ 0.15, and therefore assume the pressure drop to be constant for the duration of the experiment experiment.

\subsection{Experimental protocols}
For experiments, we added stock to surfactant solution at a strongly supramicellar target concentration of \SI{7.5}{\wtpc} TTAB, at a ratio of 1:60 between stock and swimming solutions and pipetted them into the channel structure. We started recording videomicroscopy once a droplet had navigated into the region of interest.
\subsection*{Analysis}
For the PIV analysis of the flow field, we added tracer colloids (FluoSpheres\textsuperscript{TM} F8812, 0.5 $\mu$m, red (580/605), Thermo Fisher) to the swimming medium at a volume fraction of  $10^{-3}$.

We mounted the microfluidic chips on an inverted  microscope (Olympus IX81), with the focal plane set to the middle of the channel structure. The electroosmotic flow (EOF) was recorded under 40x magnification, the droplet motion under magnifications varying from 10x to 40x.
For the droplet experiments, fluorescence and bright field microscopy were combined.

We measured the EOF and droplet flow field (PIV) using a high-speed camera (Phantom v311) at 100 fps. Since recording was time limited to \SI{8.2}{\second} owing to camera buffer limits, we used a Canon EOS 600D DSLR camera for long-time droplet tracking, with an image size of 1920 px × 1080 px and a framerate of 25 fps.

\subsection*{Particle Image Velocimetry (PIV)}
We analyzed the electroosmotic and droplet-driven flow velocities by Particle Image Velocimetry (PIV)~\cite{adrian2011_particle} on videomicroscopy data recorded at 100 fps, using the open-source MATLAB
module PIVlab~\cite{thielicke2014_pivlab}. The flow field grid spacing was 5x5 px$^2$ at a magnification of 40x (Interrogation area: Pass 1: 24, pass 2: 16 and pass 3: 10), yielding a spatial resolution of approximately 2.5x\SI{2.5}{\um^2}. Wall slip below this length scale could not be resolved.
\subsection*{Droplet tracking}
We tracked droplets by custom Python scripts using the openCV library \cite{bradski2000_opencv}. Preprocessing consisted of greyscale conversion, background correction, a Gaussian blur to filter out the tracer colloids and a final binarization step. 
We extracted droplet centroids, and from these trajectories via a frame-by-frame nearest neighbor algorithm~\cite{crocker1996_methods}.
\subsection{Electroosmotic flow calibration}
\begin{figure}[h!]
    \centering
    \includegraphics[width=0.9\columnwidth]{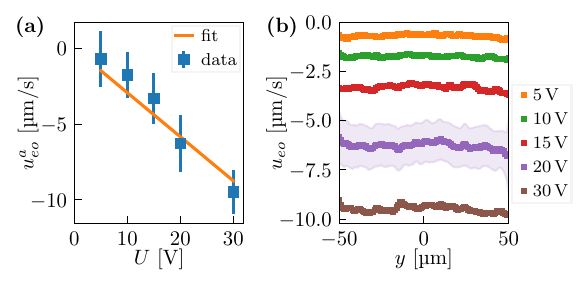}
    \caption{Electroosmotic flow calibration for TTAB solution depending on the external voltage. Data from PIV in an experiment without droplets. Mean flow speeds (a) and flow speed profiles across the channel width at different voltages (b).}
    \label{sifig:eocalibration}
\end{figure}
To calibrate the electroosmotic mobility, we recorded the plug flow of an aqueous solution of \SI{7.5}{\wtpc} TTAB in microchannels via PIV as described above, under increasing applied voltage $U$. We calculated the electroosmotic mobility, $\mathcal{M}_\text{eo}\approx \SI{-4.4e-9}{\m^2/\volt\second}$, via a zero intercept linear regression fit to $u_\text{eo}=\mathcal{M}_\text{eo}\cdot U/L$, with $L=\SI{1.5}{\cm}$

\subsection*{Reverse Hopf bifurcation during electro-rheotaxis}

\end{document}